\documentclass[3p,times,procedia]{elsarticle}
\usepackage{ecrc}
\usepackage[bookmarks=false]{hyperref}
    \hypersetup{colorlinks,
      linkcolor=blue,
      citecolor=blue,
      urlcolor=blue}
\usepackage{float}      
\usepackage{amsmath}
\usepackage{makecell}
\usepackage{multirow}

\volume{00}

\firstpage{1}

\journalname{Procedia Computer Science}

\runauth{T. Liu et al.}

\jid{procs}

\usepackage{amssymb}

\usepackage[figuresright]{rotating}

\usepackage{subfigure}

\begin{document}
\begin{frontmatter}



\dochead{Proceedings of International Conference on Biomimetic, Intelligence and Robots}

\title{Landmark Detection using Transformer Toward Robot-assisted Nasal Airway Intubation}


\author[a]{Tianhang Liu}
\author[a]{Hechen Li}
\author[a]{Long Bai}
\author[a,b,c]{Yanan Wu}
\author[a]{An Wang}
\author[d]{Mobarakol Islam}
\author[a,e,f,g]{Hongliang Ren\corref{cor1}}

\address[a]{Department of Electronic Engineering, The Chinese University of Hong Kong, Hong Kong, China}
\address[b]{College of Medicine and Biological Information Engineering, Northeastern University, Shenyang, China}
\address[c]{Key Laboratory of Intelligent Computing in Medical Image, Ministry of Education, Northeastern University, Shenyang, China}
\address[d]{Wellcome/EPSRC Centre for Interventional and Surgical Sciences (WEISS), University College London, London, UK}
\address[e]{Department of Biomedical Engineering, National University of Singapore, Singapore}
\address[f]{Shun Hing Institute of Advanced Engineering, The Chinese University of Hong Kong, Hong Kong, China}
\address[g]{Shenzhen Research Institute, The Chinese University of Hong Kong, Shenzhen, China}
\begin{abstract}
Robot-assisted airway intubation application needs high accuracy in locating targets and organs. Two vital landmarks, nostrils and glottis, can be detected during the intubation to accommodate the stages of nasal intubation. Automated landmark detection can provide accurate localization and quantitative evaluation. The Detection Transformer (DeTR) leads object detectors to a new paradigm with long-range dependence. However, current DeTR requires long iterations to converge, and does not perform well in detecting small objects. This paper proposes a transformer-based landmark detection solution with deformable DeTR and the semantic-aligned-matching module for detecting landmarks in robot-assisted intubation. The semantics aligner can effectively align the semantics of object queries and image features in the same embedding space using the most discriminative features. To evaluate the performance of our solution, we utilize a publicly accessible glottis dataset and automatically annotate a nostril detection dataset. The experimental results demonstrate our competitive performance in detection accuracy. Our code can be accessible at \href{https://github.com/ConorLTH/airway_intubation_landmarks_detection}{https://github.com/ConorLTH/airway\_intubation\_landmarks\_detection}.

\end{abstract}

\cortext[cor1]{Corresponding author}
\end{frontmatter}

\email{hlren@ee.cuhk.edu.hk}



\section{Introduction}
Object detection is critical in computer vision, necessitating accurate recognition and localization of objects within frames. Deep learning methods have spurred the creation of numerous models that can detect objects efficiently and effectively. Recently, Detection Transformer (DeTR)~\cite{detr} was introduced, leading to novel design methodologies. Before DeTR, human-designed elements and detection processes heavily impacted detector performance. However, DeTR enables fully end-to-end detection frameworks. Nevertheless, DeTR's performance is limited by two issues: low accuracy for detecting small objects and slow convergence rates.

To address these limitations, researchers developed Deformable DeTR~\cite{zhu2020deformable}. Inspired by Deformable Convolution~\cite{dai2017deformable,zhu2020deformable}, this model provides better performance on small objects. Researchers noted that DeTR's slow convergence was primarily due to challenges in matching object queries with features, particularly during cross-attention in the decoder. To address this challenge, they suggested methods that would help object queries match with objects' features. For instance, Conditional DeTR~\cite{meng2021conditional} splits the object query into query content and query spatial embedding to search relevant areas based on objects' appearances and locate extremity areas. Spatially modulated co-attention (SMCA) DeTR~\cite{smca} introduces a Gaussian spatial map that forces cross-attention to focus on specific local spatial areas. Semantic-Aligned-Matching (SAM) DeTR~\cite{sam} designs a flexible module that implements semantics-align to simplify the matching of objects' features and object queries.

The development of computer vision has propelled its application in many medical fields, such as image pre-processing~\cite{bai2023llcaps,chen2022exploiting,chen2022lsvc}, medical diagnosis~\cite{bai2022transformer,che2023image,che2022learning}, robot-assisted surgery and intubation~\cite{bai2023surgical,wang2023domain,zhang2022deep}. One such application is airway intubation, which may require nasal intubation for semiconscious or awake patients when a laryngoscope is not an option but poses challenges due to insufficient space. Intubation during medical procedures can pose significant challenges. To overcome these challenges, some researchers have implemented convolutional neural networks (CNNs) to classify and detect intubation difficulties~\cite{wu2022two,zhao2022cot}. For example, Hayasaka \emph{et al.}~\cite{hayasaka2021creation} developed an AI model that uses the patient's facial image to classify intubation difficulty. Similarly, Aguilar \emph{et al.}~\cite{aguilar2020detection} proposed a mobile app that employs a CNN to detect a difficult airway. Specifically, this work defines two classes for a challenging airway based on the Mallampati score. Deep learning in difficult airway detection has shown promise in facilitating intubation procedures. However, there are several areas for improvement. For instance, existing models focus solely on detecting difficult airways, without additional information on intubation procedures, which can limit their practicality. Therefore, a more comprehensive approach to difficult airway detection is necessary. However, the need for medical professionals to label these datasets manually has limited further progress. Moreover, creating datasets for specific applications based on current open-source datasets and automatically generating annotations can overcome this limitation. 

To overcome these limitations, we present a novel approach for achieving comprehensive landmark detection to enhance the efficiency of intubation procedures. Landmarks serve as essential indicators during the intubation process, providing valuable information about its progress. The specific landmarks to be detected may vary at different stages of the intubation process. For example, during the initial phase, when the tube is external to the human body, detecting the position of the nostril becomes essential. Subsequently, as the tube moves inside the nasal passage, detecting the glottis becomes crucial. However, some landmarks, such as nostrils, are relatively small objects, presenting a challenge for accurate detection by methods like DeTR. In this work, we make the following contributions to address these challenges:

\begin{itemize}
    \item We propose a comprehensive solution to facilitate the successful execution of robot-assisted intubation. Our method demonstrates the capability to effectively detect pertinent anatomical landmarks, thereby furnishing critical information required for precise intubation procedures.
    \item We integrate the semantic aligner and deformable attention module, into the detection transformer framework for landmark detection tasks. This integration enhances the model's ability to identify and localize landmarks throughout the intubation process accurately.
    \item Additionally, we devise an automatic annotation methodology for generating detection bounding boxes for a nostril dataset. Furthermore, through extensive experimentation, we provide compelling evidence showcasing the superior performance of our proposed solution in accurately detecting landmarks during intubation. Our transformer-based detection solution achieves outstanding performance with only 24-epoch training.
\end{itemize}

\section{Detection Model}
In this section, we will discuss the DeTR model, its current challenges, and the process of adapting DeTR for the practical application of airway landmark detection.

\subsection{DeTR and its Challenges}
DeTR is a state-of-the-art object detection model that treats object detection as a set prediction problem. It uses learnable object queries to form relations with extracted features using the transformer-based encoder and decoder modules. In the encoder, self-attention modules encode the extracted image features. In the decoder, two attention mechanisms - self-attention modules and cross-attention modules - are utilized to enable information exchange between object queries and encoded features. DeTR faces two challenges. Firstly, the computation complexity of the attention mechanism increases with the spatial size of the input, making it difficult to implement multi-level feature maps and improve small object detection accuracy~\cite{zhu2020deformable}. Secondly, the random initialization of object queries makes them pair equally with all spatial locations of the image, rather than specific regions for detecting objects~\cite{gao2021fast,meng2021conditional,zhu2020deformable}. This prevents training object queries from focusing on specific regions. Therefore, long training iterations are necessary for DeTR to perform well.

To address the computation complexity challenge, Deformable DeTR was proposed. It combines deformable convolution with the attention mechanism to create Multi-scale Deformable Attention~\cite{zhu2020deformable}. This novel approach can aggregate features from multi-level feature maps and decrease the computation complexity of the original attention mechanism used in DeTR. The deformable attention only samples key positions on the feature map to act as queries, rather than considering all possible spatial locations.

Another variant, SAM-DeTR, proposes the Semantic Aligner Module to accelerate DeTR's convergence. To ensure semantic alignment, it aligns object queries and features within uniform embedding spaces. SAM uses learnable reference boxes to guide object queries to align with the semantics of encoded features. This process involves three sub-operations: extracting region features with reference boxes using ROIAlign; resampling salient points with $M$ points for each region to form new object queries and their corresponding positional embeddings; and evaluating re-weighting coefficients with current object queries to combine messages of old and new object queries. The operations related to object queries are formulated using the sigmoid function.

\subsection{Landmark Detection Solution}

\begin{figure}[ht]
    \centering
    \includegraphics[width=0.8\linewidth, trim=0 0 0 0]{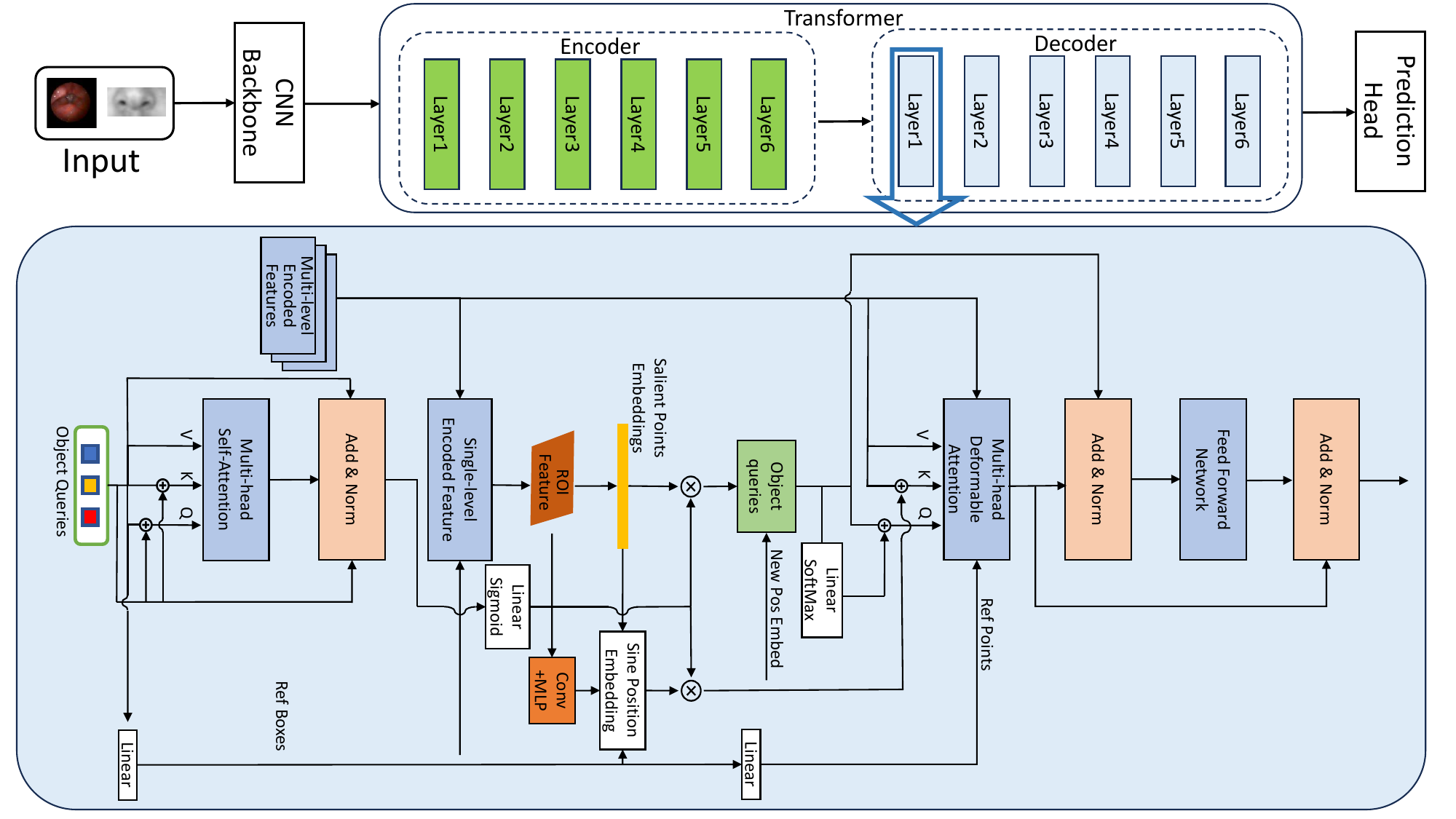}
    \caption{The workflow of our proposed solution with deformable attention and semantic aligner.}
    \label{fig:workflow}
\end{figure}

We propose integrating the semantic aligner module into Deformable DeTR, which combines deformable convolution with multi-scale deformable attention. To achieve this integration, we focus on adapting the decoder in Deformable DeTR to accommodate the SAM module. Specifically, we model the object queries and their positional embeddings as $query$, $query_{pos} \in \mathbb{R}^{N,256}$, following Deformable DeTR's implementation, instead of directly modeling the object query of $\mathbb{R}^{N,4}$ used in the SAM-DeTR for modeling reference boxes where N represents the number of the object queries and 4 and 256 are hyperparameters for the channel of queries. We also use four feature levels in multi-scale deformable attention modules, consistent with the implementation in Deformable DeTR. In contrast, the SAM module of each layer in the decoder only uses encoded features from a single level to realize semantic alignment between object queries and encoded features. To achieve this, different decoder layers take encoded features from different levels as input.

The reference boxes used in the SAM module are obtained via a learnable linear projection over query position embeddings, while the reference points used in the cross-attention modules of the decoder are obtained from a learnable linear projection over reference boxes. To avoid inappropriate reference points being chosen, we do not directly regard the center points as the reference points, even though the reference boxes are represented as coordinates of the center points and the height and width of the bounding boxes.

To ensure interoperability between the cross-attention and SAM modules, we set the hidden dimension of cross-attention to $256 \times 8$. Here, 256 is the dimension of the object queries, and 8 is the number of heads in multi-head attention. In the SAM module, each object query is assigned a reference box. During resampling, each reference box is resampled with $M$ salient points. $M$ representing the number of heads in multi-head attention (typically 8). Thus, after resampling, the dimension of the query changes to $Q^{new} \in \mathbb{R}^{N \times M \times 256}$. Before sending $Q^{new}$ into the cross-attention module, it is reshaped to $[N, M \times 256]$. Therefore, our approach integrates the SAM module into Deformable DeTR, providing a more effective and efficient landmark detection model. The overview architecture of our solution can be found in Fig.~\ref{fig:workflow}.

\section{Experiments}
\subsection{Datasets}

In the context of this research, we utilized two distinct datasets: one for nostril detection and another for glottis detection, with the aim of enabling airway intubation landmark detection.

The nostril dataset is derived from the publicly available BioID dataset~\cite{jesorsky2001robust}, which was initially collected and annotated for human face keypoint detection. The dataset consists of 1521 grayscale images, all standardized to a resolution of 384x286 pixels. We manually use key point annotations of nostrils provided by the BioID dataset as the center points and expand the height and width to form bounding box annotations. To denote the nostril locations, we adopt the bounding box annotation format following the COCO format~\cite{lin2014microsoft}.

Similarly, the glottis dataset is based on the publicly available BAGALS dataset~\cite{gomez2020bagls}, initially designed for glottis segmentation. Collaboratively compiled by seven institutions, the BAGALS dataset comprises endoscope videos, detailed segmentation annotations, and frame-level information. From this collection, we extracted a subset of explicit annotations from nasal endoscopic videos, resulting in 881 images for glottis detection. As the videos originate from diverse sources, the frames within the subset exhibit varying resolutions. To form corresponding bounding box annotations, we use OpenCV to capture the contours of the segmentation annotations of BAGALS and annotate with their maximum and minimum coordinate values. We follow the COCO format for annotating the glottis dataset to maintain consistency.

\subsection{Implementation Details}
In our experiments, the object detection models are implemented using MMdetection~\cite{chen2019mmdetection}, a freely available toolbox for PyTorch that supports various deep learning-based detectors. All models are trained from scratch and evaluated using the abovementioned two datasets. The nostril dataset is partitioned into 1021 frames for training, 185 frames for validation, and 315 frames for testing. Similarly, the glottis dataset is divided into 377 frames for training, 88 frames for validation, and 416 frames for testing. For feature extraction, we opt for the ResNet50 backbone~\cite{he2016deep}. Besides, we employed the Adam optimizer~\cite{kingma2014adam} with a learning rate of $1 \times 10^{-5}$ for the backbone and $1 \times 10^{-4}$ for the detectors. We ensure standardized implementation and evaluation across all models by utilizing MMdetection and adhering to this experimental setup, facilitating reliable comparisons and analysis. We set the number of epochs to $24$ to observe the performance under limited training iterations.

\subsection{Experimental Results}
\begin{table}[ht]
\centering
\caption{Comparison experiments of our SAM Deformable DeTR against baseline models on the glottis dataset.}
\label{tab:B}
\begin{tabular}{cccc}
\noalign{\smallskip}\hline
Method& mAP & mAP@0.5 & mAP@0.75 \\\hline
FasterRCNN~\cite{ren2015faster}& $0.129$& $0.366$& $0.054$ \\
YOLOv3~\cite{redmon2018yolov3}& $0.254$& $\textbf{0.733}$& $0.084$ \\
FCOS~\cite{tian2019fcos}& $0.042$& $0.188$& $0.002$ \\
Centernet~\cite{zhou2019centernet}& $0.037$& $0.153$& $0.006$ \\
Deformable DeTR~\cite{zhu2020deformable}& $0.128$& $0.482$& $0.013$ \\
SAM-DeTR~\cite{sam}& $0.015$& $0.064$& $0.003$ \\
$\textbf{Ours}$& $\textbf{0.282}$& $0.661$& $\textbf{0.270}$ \\\hline
\end{tabular}
\end{table}

\begin{table}[ht]
\centering
\caption{Comparison experiments of our SAM Deformable DeTR against baseline models on the nostril dataset.}
\label{tab:N}
\begin{tabular}{cccc}
\noalign{\smallskip}\hline
Method& mAP& mAP@0.5& mAP@0.75 \\
\hline
YOLOv3~\cite{redmon2018yolov3}&  $0.311$& $0.855$& $0.133$ \\
FCOS~\cite{tian2019fcos}&  $0.084$& $0.267$& $0.027$ \\
Centernet~\cite{zhou2019centernet}&  $0.171$& $0.673$& $0.017$ \\
$\textbf{Ours}$& $\textbf{0.325}$& $\textbf{0.865}$& $\textbf{0.142}$ \\
\hline
\end{tabular}
\end{table}

\begin{figure}[t]
\centering 
\subfigure[mAP on glottis dataset]{
\label{mapB}
\includegraphics[width=8cm]{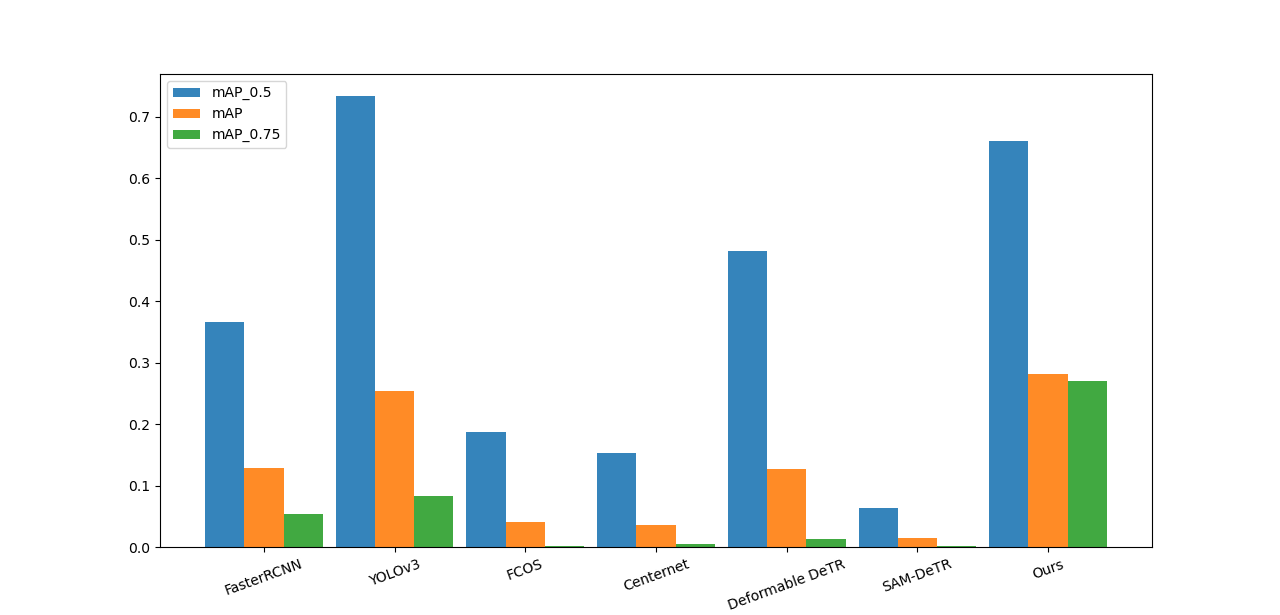}}\subfigure[mAP on nostril dataset]{
\label{mapN}
\includegraphics[width=8cm]{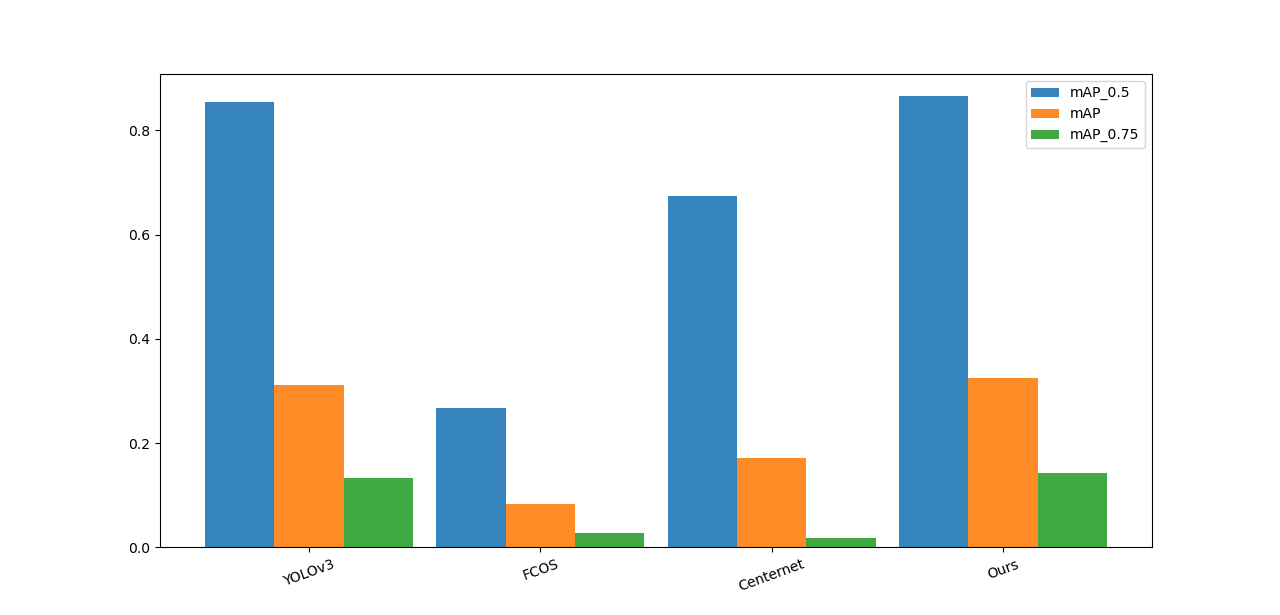}}
\caption{The visualization of the mAP results from the comparison experiments on the glottis and nostril dataset.}
\label{map}
\end{figure}

\begin{figure}
    \centering
    \includegraphics[width=\linewidth, trim=0 60 150 0]{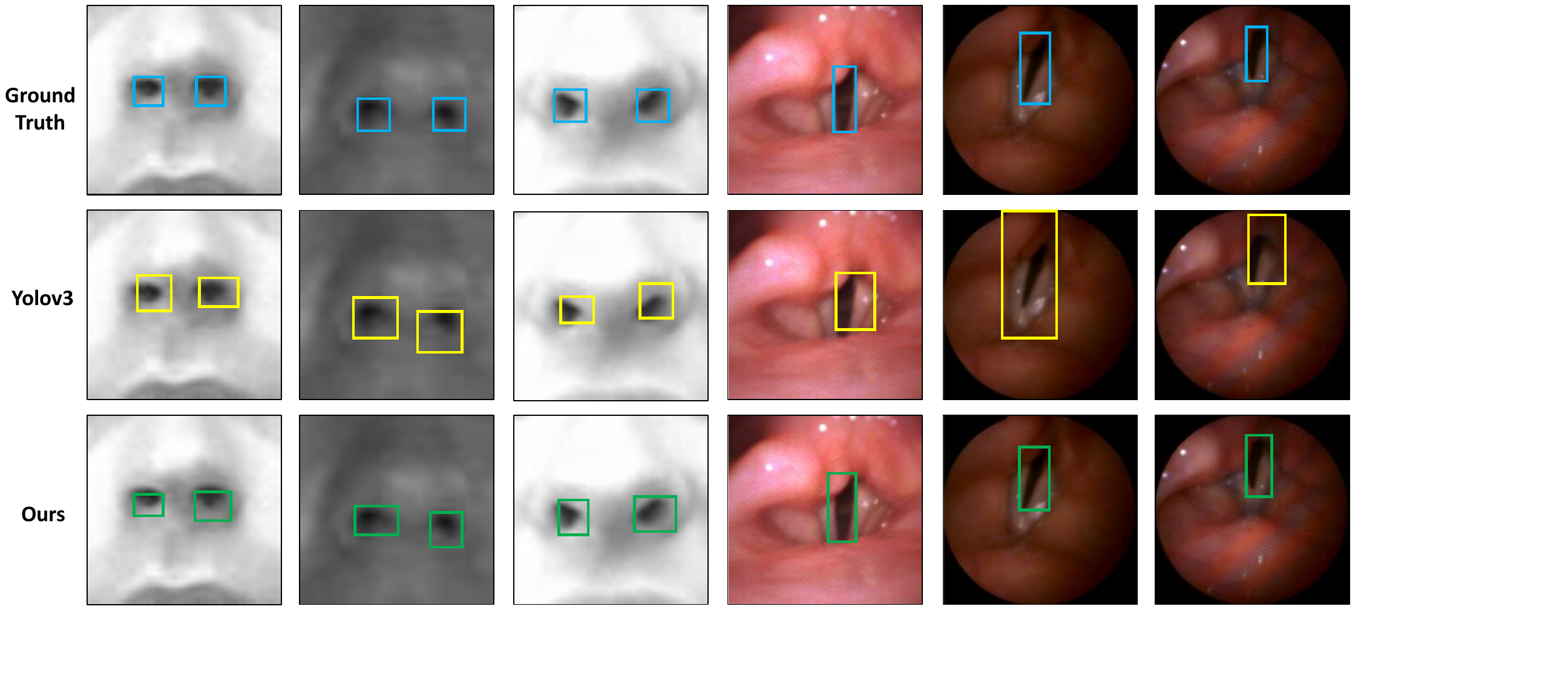}
    \caption{The qualitative comparison of the experimental results on the nostril (column 1-3) and glottis (column 4-6) detection task.}
    \label{fig:vis}
\end{figure}

Tables~\ref{tab:B} and \ref{tab:N} present the results obtained from various models trained on two distinct datasets using a consistent 24-epoch training scheme. The baseline models consist of both CNN-based and Transformer-based architectures. The performance evaluation adheres to the standard COCO metric. We also visualize the qualitative results in Fig.~\ref{fig:vis}.

As depicted in Table \ref{tab:B} and Figure \ref{mapB}, the test results on the glottis dataset reveal the advantages of combining the semantics aligner module with deformable attention, as evident from the performance of SAM Deformable DeTR and other baseline models trained from scratch. Notably, SAM-DeTR exhibits inferior results on the glottis dataset. This can be attributed to the inherent limitations of DeTR, despite the inclusion of a plug-and-play module in SAM-DeTR to expedite its convergence. SAM-DeTR encounters similar challenges as DeTR when trained on datasets with limited capacity. In contrast, the performance of Deformable DeTR surpasses that of SAM-DeTR, although it falls short compared to certain CNN-based detectors. These findings suggest that integrating the semantic aligner module into Deformable DeTR has the potential to enhance the model's capabilities, an observation supported by the results of SAM Deformable DeTR. Specifically, the mAP increment demonstrates an 18-fold increase compared to SAM-DeTR and a 2-fold increase compared to Deformable DeTR.

In Table \ref{tab:N} and Figure \ref{mapN}, the performance of SAM Deformable DeTR on the nostril dataset achieves the highest capability. The Deformable DeTR and SAM-DeTR results are omitted from Table \ref{tab:N} since their mAP results are 0.000. This suggests that these models were not fully trained on the nostril dataset with limited training iterations. The nostril dataset lacks distinct appearances, and the features of the nostril are strongly correlated with the fixed locations and surrounding facial organs. SAM Deformable DeTR effectively leverages multi-scale features and spatial information in object detection, contributing to its superior performance on this dataset.

\section{Conclusion}
In this research work, we employ SAM Deformable DeTR, an improved version of the Deformable DeTR framework integrated with the semantics aligner module inspired by SAM-DeTR. Our primary objective is to evaluate the performance of our solution and demonstrate its effectiveness in detecting airway intubation landmarks for medical applications. Comprehensive experiments were conducted using two public datasets, encompassing various scenarios encountered in airway intubation procedures. The experimental results indicate that SAM Deformable DeTR surpasses both the original SAM-DeTR framework and the standard Deformable DeTR model regarding detection accuracy. The outcomes of our study demonstrate the significant advancements of our solution in the domain of airway intubation landmark detection. The proposed framework exhibits superior performance compared to existing methods, showcasing its potential for improving medical interventions and facilitating accurate and efficient airway-related procedures.

\section*{Acknowledgements}
This work was supported by Hong Kong Research Grants Council (RGC) Research Impact Fund (RIF) R4020-22, Collaborative Research Fund (CRF C4026-21GF, CRF C4063-18G), General Research Fund (GRF 14203323),  NSFC/RGC Joint Research Scheme N\_CUHK420/22, GRS \#3110167; Shenzhen-Hong Kong-Macau Technology Research Programme (Type C) STIC Grant SGDX20210823103535014 (202108233000303); Guangdong Basic and Applied Basic Research Foundation (GBABF) \#2021B1515120035; Shun Hing Institute of Advanced Engineering (SHIAE Project BME-p1-21) at The Chinese University of Hong Kong (CUHK).








\bibliographystyle{cas-model2-names}

\bibliography{ref}




\clearpage

\normalMode







\end{document}